\documentclass[submission,copyright,creativecommons]{eptcs}
\providecommand{\event}{ACT 2020}
\title{
  Reverse Derivative Ascent: \\
  A Categorical Approach to Learning Boolean Circuits
}
\author{
  Paul Wilson
  \institute{University College London}
  \institute{University of Southampton}
  \email{paul@statusfailed.com}
  \and
  Fabio Zanasi
  \institute{University College London}
  \email{f.zanasi@ucl.ac.uk}
}
\def\authorrunning{P.W. Wilson, F. Zanasi}
\def\titlerunning{Reverse Derivative Ascent}
\date{\today}

\usepackage[utf8]{inputenc}
\usepackage{booktabs}

\usepackage{bbm} 
\usepackage{calrsfs}
\usepackage{multicol}
\usepackage{xcolor}
\DeclareMathAlphabet{\pazocal}{OMS}{zplm}{m}{n}

\usepackage{float}
\usepackage{amsmath}
\usepackage{amsthm}
\usepackage{amssymb}
\usepackage{thmtools}
\usepackage{hyperref}
\usepackage{enumitem}

\usepackage{algorithm}

\usepackage[noend]{algpseudocode}

\usepackage{stmaryrd} 
\usepackage{tikzit}   

\tikzstyle{conode}=[fill=black, draw=black, shape=circle]
\tikzstyle{node}=[fill=white, draw=black, shape=circle]
\tikzstyle{box}=[fill=white, draw=black, shape=rectangle]
\tikzstyle{medium box}=[fill=white, draw=black, shape=rectangle, minimum width=0.75cm, minimum height=1.2cm]
\tikzstyle{boundary box}=[fill=white, draw=black, shape=rectangle, minimum width=0.75cm, minimum height=2.5cm]
\tikzstyle{green}=[fill={rgb,255: red,0; green,215; blue,0}, draw={rgb,255: red,0; green,215; blue,0}, shape=circle]
\tikzstyle{red}=[fill=red, draw=red, shape=circle]
\tikzstyle{antipode}=[fill=red, draw=red, shape=rectangle]

\tikzstyle{separator}=[-, draw={rgb,255: red,128; green,128; blue,128}, dashed]
\tikzstyle{arrow}=[->]
\tikzstyle{dottedarrow}=[->, draw={rgb,255: red,128; green,128; blue,128}, dashed]

\theoremstyle{definition}
\newtheorem{theorem}{Theorem}
\newtheorem{lemma}[theorem]{Lemma}
\newtheorem{proposition}[theorem]{Proposition}
\newtheorem{corollary}[theorem]{Corollary}
\newtheorem{definition}[theorem]{Definition}

\newtheorem{remark}[theorem]{Remark}
\newtheorem{example}[theorem]{Example}

\newcommand{\from}{:}
\newcommand{\RDA}{Reverse Derivative Ascent}

\newcommand{\catname}[1]{\ensuremath{\mathrm{\mathbf{#1}}}}
\newcommand{\functor}[1]{\ensuremath{\mathrm{\mathbf{#1}}}}
\newcommand{\textdef}[1]{\emph{#1}}

\newcommand{\Z}{\ensuremath{\mathbb{Z}_2}}
\newcommand{\ToPoly}[1]{\ensuremath{\llbracket {#1} \rrbracket^{\mathbb{P}}}}
\newcommand{\ToBoolFun}[1]{\ensuremath{\llbracket {#1} \rrbracket^\mathbb{B}}}
\newcommand{\ToCanonical}[1]{\ensuremath{\functor{C}({#1})}}

\newcommand{\BoolFun}{\catname{BoolFun}}

\newcommand{\BoolCirc}{\catname{BoolCirc}}
\newcommand{\PolyCirc}{\catname{PolyCirc}}
\newcommand{\PolyZ}{\ensuremath{\catname{Poly}_{\Z}}}

\newcommand{\inlinefig}[1]{
  \resizebox{!}{\fontcharht\font`\B}{\input{#1.tikz}}
}

\newcommand{\biginlinefig}[2]{
  \resizebox{!}{#2}{\input{#1.tikz}}
}

\newcommand{\Partial}[2]{\ensuremath{D_{#1}[#2]}}

\newcommand{\FZ}{\textbf{0}}

\newcommand{\gcopy}[0]{\inlinefig{copy}}
\newcommand{\gand}[0]{\inlinefig{and}}
\newcommand{\gadd}[0]{\inlinefig{add}}
\newcommand{\gdiscard}[0]{\inlinefig{discard}}
\newcommand{\gzero}[0]{\inlinefig{zero}}
\newcommand{\gone}[0]{\inlinefig{one}}

\newcommand{\gidentity}[0]{\inlinefig{identity}}

\newcommand{\setname}[1]{\ensuremath{\mathcal{#1}}}

\newcommand{\Axioms}{\ensuremath{\setname{A}}}

\newcommand{\procname}[1]{\ensuremath{\mathrm{\texttt{#1}}}}
\newcommand{\rdaStep}{\procname{rdaStep}}
\newcommand{\rda}{\procname{rda}}

\newcommand{\pow}[1]{\ensuremath{\mathrm{pow}(#1)}}

\newcommand{\eqmod}[1]{\ensuremath{\stackrel{#1}{=}}}
\newcommand{\R}{\ensuremath{\widetilde{R}}}

\begin{document}

\maketitle

\begin{abstract}
  We introduce \emph{Reverse Derivative Ascent}: a categorical analogue of gradient based
  methods for machine learning.
  Our algorithm is defined at the level of so-called
  \emph{reverse differential categories}.
  It can be used to learn the parameters of models which are expressed as
  morphisms of such categories.
  Our motivating example is boolean circuits: we show
  how our algorithm can be applied to such circuits by using the theory of
  reverse differential categories.
  Note our methodology allows us to learn
  the parameters of boolean circuits \textit{directly}, in
  contrast to existing binarised neural network approaches.
  Moreover, we demonstrate its empirical value by giving experimental
  results on benchmark machine learning datasets.
\end{abstract}

\section{Introduction}

Computation of the reverse derivative is a critical part of gradient-based
machine learning methods (see e.g.~\cite{ruder_overview_2017} for an overview).
In essence, the reverse derivative tells us how to update the parameters of a model,
given a prediction error.
This update procedure is at the core of many optimisation methods, such as
\textit{stochastic gradient descent} \cite[2.2]{ruder_overview_2017} --used for training deep neural
networks.

Now, the model class of choice is typically neural networks, which can be
considered as the smooth maps $\mathbb{R}^a \to \mathbb{R}^b$. A natural
question to ask is whether these gradient based methods could be generalised to
other settings. In this paper, we focus on boolean
functions - maps $\Z^a \to \Z^b$, and their operational counterpart, boolean circuits.

This setting has real practical value. Larger neural network models typically require expensive and power-hungry GPGPU hardware to train and
run \cite{courbariaux_binaryconnect_nodate} \cite{raina_large-scale_2009}.
Performance can be improved via \emph{binarisation}: the extraction from a
trained neural network of a boolean circuit, which provides a better optimised
representation of the same model.

The usual pattern in these approaches (see e.g. BinaryConnect
\cite{courbariaux_binaryconnect_nodate} and LUTNet \cite{wang_lutnet:_2019}) is
to perform the training aspect exclusively on the `real-valued' side.
However, training schemes for binarised models such as boolean circuits are
typically more efficient~\cite{hubara_binarized_nodate}. It is thus natural to
ask: ``why not learn the parameters of boolean circuits directly?''

\RDA{}, the algorithm that we introduce in this paper, originates from this question: instead of training a neural network and then
extracting a boolean circuit, we begin with a boolean circuit, and learn its parameters
directly.

In defining and analysing the algorithm, we take a categorical approach. Our methodology relies on the abstract framework provided by \emph{reverse differential categories} \cite{cockett_reverse_2019}, which axiomatises the concept of reverse derivative operator. We proceed in three steps:
\begin{enumerate}
  \item
    We give a syntactic presentation in terms of string diagrams for the
    reverse differential operator of polynomials, and a safety condition
    specifying when we can apply this operator to boolean circuits.
	\item We define \RDA{} as `gradient'-based algorithm working for arbitrary morphisms of reverse differential categories.
	\item We can then apply \RDA{} to boolean circuits, and demonstrate its empirical value by giving experimental
    results for benchmark datasets.
\end{enumerate}
The categorical setting brings two main advantages. First, by exploiting the
presentation of boolean circuits as an axiomatic theory of string
diagrams~\cite{lafont_towards_2003}, we are able to define a suitable reverse
derivative operator \emph{compositionally}, by induction on the circuit syntax.
Second, because our definition of \RDA{} is phrased at the general level of
reverse differential categories, it paves the way for the application to model
classes other than boolean circuits, which we leave for future work.

The rest of the paper is structured as follows.
We begin with necessary background in Section \ref{section:background}, defining
(parametrised) boolean functions and circuits.
In Section \ref{section:graphical-boolean-reverse-derivative} we define our
graphical operator for boolean circuits,
give a safety condition for its application,
and show that it is consistent with the reverse differential combinator of
polynomials.
We include additional material relevant to
Section \ref{section:graphical-boolean-reverse-derivative} in
Appendix \ref{appendix:polynomials}.
In Section \ref{section:reverse-derivative-ascent}, we describe the \RDA{}
algorithm and provide a Haskell library to run our algorithm on boolean
circuits\footnote{
  Note that our implementation represents circuits in terms of the
  corresponding boolean functions:
  this is just a presentation choice, because the category of boolean functions
  and boolean circuits are isomorphic.
  We refer to Section \ref{subsubsection:implementation-details} for a more
  comprehensive discussion on the implementation.
}. Also, we give empirical evidence that it is able to learn functions from data.
We conclude the paper with a discussion of future work in
Section \ref{section:discussion}.

\section{Background: Boolean Functions, Circuits, and Polynomials}
\label{section:background}



We first recall the basics of boolean functions.

\begin{definition}
  A \textdef{boolean function} is a map $f \from \Z^a \to \Z^b$.
  We also say that a map $g \from \Z^{p + a} \to \Z^b$ is a
  \textdef{parametrised boolean function} with $p$ parameters,
  $a$ inputs, and $b$ outputs.
  We denote by \BoolFun{} the symmetric strict monoidal category whose objects
  are natural numbers with addition as tensor product,
  and whose morphisms are boolean functions.
  The monoidal product of this category is in fact the cartesian product, making
  this category cartesian.
\end{definition}

  While \textit{parametrised} boolean functions are of course exactly the boolean
  functions, by distinguishing the $p$ parameters from the $a$ inputs
  we mean to declare our intent:
  our goal is to learn a map $\Z^a \to \Z^b$ which approximates a given dataset
  of input/output
  \textit{examples} of the type $(x, y) \in (\Z^a, \Z^b)$.
  To do this, we must choose a particular \textit{model}: a function
  $f \from \Z^{p + a} \to \Z^b$.
  We then use our machine learning algorithm to search for a set of parameters
  $\theta \in \Z^p$
  such that $f(\theta, -) \from \Z^a \to \Z^b$ approximates the dataset well.

\begin{example}
  \label{example:eval-bool-fun}
  Suppose we wish to learn a boolean function $\Z \to \Z$
  with no prior knowledge of the dataset.
  One choice of model is the function
  $\mathrm{\texttt{eval}} \from \Z^{2 + 1} \to \Z$,
  defined as
  $\mathrm{\texttt{eval}}(\theta, x) \mapsto \theta_x$.
  That is, the function of $2$ parameters which maps the data bit $x$ to $\theta_0$
  if $x = 0$, and $\theta_1$ if $x = 1$.
  In this case, our parameters represent a \textit{truth table}: i.e.,
  they extensionally specify the $\Z^a \to \Z^b$
  function we are learning.
\end{example}

\begin{remark}
  An interesting property of the eval model is that, because there are a finite
  number of boolean functions $\Z^a \to \Z^b$,
  the entire function space can be represented with $2^a b$ parameters.
  Clearly, this makes it suitable for only small $a$, but
  \cite{wang_lutnet:_2019} demonstrate it can be profitably used as a
  compositional building block for larger models.
\end{remark}

In order to apply \RDA{} to boolean functions, we will need
to use the reverse differential operator of a related category, which we now
introduce.

\begin{definition}
  \label{definition:polyz}
  Following \cite{cockett_reverse_2019},
  let $\PolyZ$ be the category with objects the natural numbers,
  and with morphisms $p \from a \to b$
  being
  $b$-tuples of polynomials in $a$ variables.
  That is,
  $$p = \langle p_1(\vec{x}), p_2(\vec{x}), \ldots, p_b(\vec{x}) \rangle $$
  with components $p_i(\vec{x}) \in \Z[x_1, \ldots, x_a]$, the polynomial ring
  in $a$ variables over $\Z$.
  Composition of morphisms is the composition of polynomials as in
  \cite{cockett_reverse_2019},
  where the composition
  $a \xrightarrow{p} b \xrightarrow{q} c$
  is the polynomial given by
  $q( p_1(\vec{x}), p_2(\vec{x}), ..., p_b(\vec{x}) )$
\end{definition}


In order to make the relationship between \BoolFun{} and \PolyZ{} clear,
we will now recall graphical presentations of both.
Boolean functions have a well-known graphical representation as \emph{boolean
circuits}. This correspondence can be made formal by establishing an isomorphism
between $\BoolFun{}$ and a category $\BoolCirc{}$ whose morphisms are (open)
boolean circuits, see \cite[section 4]{lafont_towards_2003}. Furthermore, the
morphisms of $\BoolCirc{}$ can be pictured as the
\emph{string diagrams}~\cite{selinger_survey_2010}
freely generated by a certain signature and equations.
Similarly, morphisms of \PolyZ{} have a graphical representation
as \textit{polynomial circuits}, which is obtained by relaxing
one of the equations of \BoolCirc{}.
As we will exploit such graphical representations in our developments, we recall
them below.

\begin{definition}
  \label{definition:boolean-circuit}
  We denote with \BoolCirc{} the
  symmetric strict monoidal category whose objects are the natural numbers and whose
  morphisms are the string diagrams freely generated by generators
  \begin{equation}
    \label{equation:circuit-generators}
    \gdiscard{} \qquad \gcopy{} \qquad \gzero{} \qquad \gadd{} \qquad \gone{} \qquad \gand{}
  \end{equation}
  and, for all morphisms $f$, equations
  \begin{equation}
    \label{equation:boolcirc-equations}
    \begin{gathered}
      \inlinefig{axioms/copy-twist} = \gcopy  \qquad
      \biginlinefig{axioms/copy-assoc-top}{0.4cm} = \biginlinefig{axioms/copy-assoc-bot}{0.4cm} \qquad
      \inlinefig{axioms/copy-unit} = \gidentity \\
      \biginlinefig{axioms/f-copy}{0.3cm} = \biginlinefig{axioms/copy-ff}{0.4cm} \qquad
      \biginlinefig{axioms/f-discard}{0.3cm} = \biginlinefig{discard}{0.3cm} \\
      \inlinefig{axioms/twist-add} = \gadd \qquad
      \biginlinefig{axioms/add-assoc-top}{0.4cm} = \biginlinefig{axioms/add-assoc-bot}{0.4cm} \qquad
      \inlinefig{axioms/add-unit} = \gidentity \\
      \inlinefig{axioms/twist-mul} = \gand \qquad
      \biginlinefig{axioms/mul-assoc-top}{0.4cm} = \biginlinefig{axioms/mul-assoc-bot}{0.4cm} \qquad
      \inlinefig{axioms/mul-unit} = \gidentity \\
      \inlinefig{axioms/copy-mul} = \gidentity \qquad
      \inlinefig{axioms/copy-add} = \gdiscard \gzero \qquad
      \biginlinefig{axioms/distrib-add-mul}{0.4cm} =
        \biginlinefig{axioms/distrib-mul-add}{0.5cm}
    \end{gathered}
  \end{equation}
  We call \textdef{circuits} the string diagrams freely
  obtained by the generators in \eqref{equation:circuit-generators} and
  quotiented by the laws of symmetric monoidal categories.
  When we say \textit{boolean} circuits, however, we mean morphisms of
  \BoolCirc{}; i.e., those circuits quotiented by equations
  \eqref{equation:boolcirc-equations}.
\end{definition}

\begin{definition}
  We denote by \PolyCirc{} the symmetric strict monoidal category whose objects
  are the natural numbers, and whose morphisms are the string diagrams freely
  generated by generators
  \eqref{equation:circuit-generators}
  and equations
  \eqref{equation:boolcirc-equations}
  minus
  $\inlinefig{axioms/copy-mul} = \gidentity{}$.
  We denote this set of axioms as $\Axioms{}$,
  and call the morphisms of this category \textit{polynomial circuits}.
\end{definition}

Note the equational theories of both Boolean and
polynomial circuits
yield that $(\gadd, \gzero)$ and $(\gand, \gone)$ form
two commutative monoids and $(\gcopy, \gdiscard)$ a commutative comonoid. In
fact, the comonoid structure makes both $\BoolCirc{}$
and $\PolyCirc{}$ into cartesian categories.

For boolean circuits, one may think operationally
of $\gadd$ as the XOR gate, $\gand$ as the AND gate, and $\gcopy$ as copy. This
intuition is at the basis of the interpretation functor $\ToBoolFun{\cdot}
\colon \BoolCirc \to \BoolFun$ of boolean circuits as boolean functions. Saying
that \eqref{equation:circuit-generators} and \eqref{equation:boolcirc-equations}
present $\BoolFun$ amounts to the following statement.

\begin{proposition}[\cite{lafont_towards_2003}]
  \label{proposition:circ-fun-iso}
	$\ToBoolFun{\cdot} \colon \BoolCirc \to \BoolFun$ is an isomorphism of symmetric monoidal categories.
\end{proposition}

\begin{corollary}\label{cor:completenessfunctions}
	$\ToBoolFun{c} = \ToBoolFun{d}$ if and only if $c \stackrel{\eqref{equation:boolcirc-equations}}{=} d$.
\end{corollary}

Similarly for \textit{polynomial} circuits, we may think of
$\gadd{}$ and $\gand{}$
respectively as the two-variable polynomials $x_1 + x_2$ and $x_1 x_2$.
Saying that generators \eqref{equation:circuit-generators} and
equations $\Axioms{}$ present $\PolyCirc$ amounts to the following statement
about the interpretation functor $\ToPoly{\cdot}$:

\begin{proposition}
  \label{prop:polyz-iso-polycirc}
  $\ToPoly{\cdot} \colon \PolyCirc{} \to \PolyZ{}$ is an isomorphism of
  symmetric monoidal categories.
\end{proposition}

\begin{corollary}
  \label{proposition:polycirc-presents-polyz}
  $\ToPoly{f} = \ToPoly{g}$ if and only if $f \eqmod{\Axioms{}} g$.
\end{corollary}

\begin{proof}
  The key idea is that hom-sets $\PolyZ{}(a, b)$ and $\PolyCirc{}(a, b)$ have
  the structure of the free module over the polynomial ring
  $\Z[x_1 \ldots x_a]$, and so there exists an isomorphism between them.  See
  \autoref{appendix:polynomials} for the full proof.
\end{proof}

\begin{remark}
  We note that the axiom
  \inlinefig{ring-property-0x-is-x}
  from \cite[Figure 40]{lafont_towards_2003} is
  redundant, and can be derived from the others.
  This is because for all $n \in \mathbb{N}$, the
  hom-sets $\BoolFun{}(n, 1)$ have ring structure where $0x = 0$ is an
  elementary property.
  We discuss this in more detail in \autoref{appendix:polynomials}.
\end{remark}

\begin{example}
  \label{example:eval-circuit}
  Continuing our example of the $\mathrm{eval} \from \Z^{2+1} \to \Z$ function
  from \autoref{example:eval-bool-fun}, we show its corresponding boolean
  circuit below, with its inputs labeled.
  Note that this circuit can be equally interpreted as a morphism
  of \PolyZ{}, namely as the single $3$-variable polynomial
  $\langle \theta_1 + (\theta_1 + \theta_2) x \rangle$.
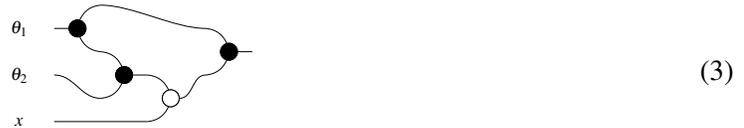
  \begin{equation}
    \label{equation:eval-2-1-circuit}
    \biginlinefig{eval-2-1-circuit}{1cm}
  \end{equation}
\end{example}

\section{
  Applying the Reverse Derivative to Boolean Circuits
}
\label{section:graphical-boolean-reverse-derivative}



In order to define our machine learning algorithm, we need a notion of
\emph{reverse derivative} for boolean circuits.
To this aim, we follow a principled approach by recalling reverse differential
categories \cite{cockett_reverse_2019}, which axiomatise the notion of a reverse
differential combinator for categorical morphisms.
  More concretely, we will translate the reverse derivative combinator of
  \PolyZ{} given in \cite{cockett_reverse_2019} to the graphical setting of
  \PolyCirc{},
  and show how we can exploit the syntactic similarity with \textit{boolean}
  circuits in order to apply it to morphisms of \BoolCirc{}.
  However, we will see that this does not make \BoolCirc{} a reverse derivative
  category, and applying the reverse derivative in this way requires a safety
  condition which we will introduce.

\begin{definition}
  (from \cite{cockett_reverse_2019}) A \textdef{reverse differential category} is a category which is
  \begin{enumerate}[label=(\roman*)]
  \item cartesian
  \item left-additive, meaning that each object $a$ is canonically equipped with a commutative monoid	 structure $(+_a \from a \times a \to a, 0_a \from I \to a)$.
  \item equipped with a \textdef{reverse differential combinator}
  which maps morphisms \inlinefig{axioms/f-morphism} to reverse derivatives
  \biginlinefig{axioms/Rf-morphism}{0.6cm}.
  obeying the axioms RD.1 through RD.7 of \cite[Section 3]{cockett_reverse_2019}.
  \end{enumerate}
\end{definition}

  Intuitively, $R[f]$ approximately computes the change in input necessary to
  achieve a given change of output for a function $f$.
  For example, suppose we have a parametrised boolean function $f \from \Z^{p+a}
  \to \Z^b$ whose predictions we denote $\hat{y} = f(\theta, x)$.
  We may have some observed data $(x, y) \in (\Z^a, \Z^b)$ that disagree with our
  predictions, i.e., where $\hat{y} \neq y$, and we wish to adjust the
  parameters of our model $\theta$ to better match our observations.
  The reverse derivative allows us to compute a \textit{change}
  \footnote{
    Note that reverse derivatives compute the changes in all inputs,
    so for a parametrised boolean circuit, this includes both parameter and data
    inputs.
  } in parameters
  $\delta_\theta$
  so that
  $f(\theta + \delta_{\theta}, x)$
  is a better prediction than
  $f(\theta, x)$
  This intuition is exactly the basis for \textit{reverse derivative ascent}, which we
  describe in \autoref{section:reverse-derivative-ascent}.

Our next goal is to show that we can apply a notion of reverse derivative to
morphisms of $\BoolCirc$.
We establish some preliminary intuition through
an example.

\begin{example}
  \label{example:boolfun-rd}
    By directly translating the definition of reverse derivative combinator for
    morphisms of \PolyZ{} to boolean functions $f \from \Z^a \to \Z^b$, we
    obtain the $a$-tuple of $(a + b)$-variable functions
    \footnote{
    We note that this definition is essentially the same as the definition
    of the reverse differential combinator
    given by
    \cite{cockett_reverse_2019} for polynomials over a semiring
    except for the definition of $D_i$,
    for which we use the partial derivatives of boolean functions given by
    \cite{martin_del_rey_boolean_2012}.
  }
  $$ \left \langle
        \sum \Partial{1}{f}(x) \cdot \delta_y ,
        \ldots ,
        \sum \Partial{a}{f}(x) \cdot \delta_y
     \right \rangle
  $$
  where we use $\cdot$ to denote pointwise multiplication of bitvectors,
  $\sum x$ to denote the sum of vector components,
  and $\Partial{i}{f}$ is the $i$th partial derivative of $f$, as defined in
  \cite{martin_del_rey_boolean_2012}: $\Partial{i}{f}(x) = f(x) + f(x + e_i)$,
  with $e_i$ the $i$th basis vector, whose entries are $0$ except for the $i$th,
  which is $1$.
\end{example}

  The above
  indeed allows us to learn the parameters of boolean circuits from data
  \footnote{
    Indeed, we expose it in our Haskell library as the function
    \texttt{RDA.ReverseDerivative.rdiffB}
  }
  but has one major flaw: efficiency.
  Computing it requires $i + 1$ evaluations of $f$,
  and in models with just a moderate number of parameters and/or where $f$ is expensive to
  compute, this quickly becomes intractable.
  However, this definition is still useful to take the reverse derivative of a
  `black box' function whose symbolic form is not known, such as a function from
  a software library.

\medskip

We will now develop a more efficient approach for boolean functions:
the key step is the introduction of $\R$, a syntactic operator defined
inductively on circuits.
However, we will see that this operation does not respect the equations
of \BoolCirc{}, and so we introduce a safety condition restricting us to those
circuits on which $\R$ is well defined.
Finally, we show that because every circuit has a safe equivalent in
\BoolCirc{},
we are able to define an operator for \BoolCirc{} which coincides with the
reverse derivative of \PolyZ{}.

\begin{definition}\label{def:Rtilde}
  For each circuit $f \from a \to b$, we define the operator $\R[f] \from a +
  b \to a$ inductively on generators \eqref{equation:circuit-generators},
  composition, and monoidal product.
  Since each generator represents a specific morphism of $\PolyZ{}$, we must
  define $\R$ on generators as

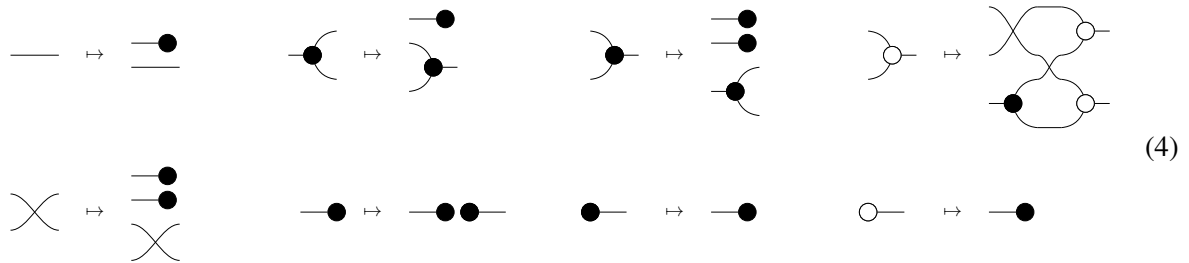
\begin{equation}
  \label{equation:r-definition-generators}
  \biginlinefig{r-definition-generators}{2cm}
\end{equation}
  Following axioms RD.5 and RD.4 of \cite[Definition 13]{cockett_reverse_2019},
  we take $\R$ on composition and monoidal product of circuits as follows:
\begin{equation}
  \label{equation:r-definition-composition}
  \biginlinefig{r-definition-composition}{1cm}.
\end{equation}
\end{definition}

Strictly speaking, since \PolyCirc{} is a cartesian category, the definition
of $\R[f \otimes g]$ is only implied by RD.4, and so we verify that
this definition indeed respects the axioms \Axioms{} of \PolyCirc{}.

\begin{lemma}
  \label{lemma:r-well-defined-on-axioms}
  $\R$ is well-defined for circuits modulo $\Axioms{}$, that is,
  $c \eqmod{\Axioms} d$ implies $\R[c] \eqmod{\Axioms} \R[d]$.
\end{lemma}

\begin{proof}
  It suffices to check the statement on $c$ and $d$ that are equal modulo a
  single axiom $\langle l,r \rangle$ in $\Axioms{}$. This means that, modulo the
  laws of symmetric monoidal categories, $c$ can be factorised as
  \biginlinefig{rewriting-factorisation-l}{0.5cm} 
  and $d$ as
  \biginlinefig{rewriting-factorisation-r}{0.5cm}.
  Thus $\tilde{R}[c] = \tilde{R}[c_L \, ; \, (id \otimes l) \, ; \, c_R]$ and
  $\tilde{R}[d] = \tilde{R}[c_L \, ; \, (id \otimes r) \, ; \, c_R]$.
  Unravelling these circuits according to Definition~\ref{def:Rtilde}, one may
  observe that in order to prove that $\tilde{R}[c] = \tilde{R}[d]$ we only need
  to check that $\tilde{R}[l] = \tilde{R}[r]$. This can be verified exhaustively
  for all $\langle l,r \rangle \in \Axioms{}$.
\end{proof}

Consequently, it is clear that this syntactic definition of $\R$ is equivalent
to the reverse derivative $R$ of \PolyCirc{} and, by isomorphism, \PolyZ{}.
However, this definition is \textit{not} compatible with boolean circuits:
although we have the axiom $\inlinefig{axioms/copy-mul} = \gidentity{}$,
we can derive $\R[\inlinefig{axioms/copy-mul}] = \inlinefig{axioms/R-copy-mul}$
and $\R[-] = \inlinefig{axioms/R-identity}$,
which are clearly not equal.
Indeed, \autoref{lemma:r-well-defined-on-axioms} highlights that this axiom is
the \textit{only} problematic one.

\medskip

To address this issue, we now introduce a condition called safety, and show that
$\R$ always respects \eqref{equation:boolcirc-equations} when applied to safe
circuits.
In essence, the following series of results give us a recipe to take the
reverse derivative of a boolean circuit, even though \BoolCirc{} does not form a
reverse derivative category.

\subsection{Safety}
\label{subsection:r-well-defined-proof}

Safety can be succinctly defined by regarding a boolean circuit
combinatorially as a directed graph.\footnote{This can be made completely formal
by interpreting string diagrams as (directed) hypergraphs with boundaries
\cite{bonchi_rewriting_2016, zanasi_rewriting_2017}, where generators form
hyperedges and wires connecting them are the nodes. In this context,
reachability between wires  (as in Definition~\ref{def:safe} below) can be
defined as the existence of a forward path between the corresponding nodes in
the hypergraphs.}

\begin{definition}\label{def:safe}
  We say a circuit $c$ is \textdef{safe} if, for every $\gand$ generator in $c$,
  the two input ports of $\gand$ are not reachable from the same input port of $c$.
\end{definition}

\begin{example}
  The circuit $\inlinefig{axioms/copy-mul}$ is not safe, whereas
  \eqref{equation:eval-2-1-circuit} is safe. The following circuit, which is
  equivalent in $\BoolCirc$ to
  \eqref{equation:eval-2-1-circuit}, is not safe.
  Indeed, both inputs of the rightmost $\gand$ generator are reachable from
  $\theta_1$ (and actually also from $\theta_2$).
  \begin{equation}
    \label{equation:unsafe-eval-circuit}
    \biginlinefig{eval-2-1-circuit-unsafe}{1cm}
  \end{equation}
\end{example}

As witnessed by \eqref{equation:eval-2-1-circuit} and \eqref{equation:unsafe-eval-circuit}, it is actually possible to show that
\begin{lemma}
  \label{prop:Rsafenormalform}
	For each boolean circuit $c$, there is a safe boolean circuit $d$ such that $c
  \eqmod{\eqref{equation:boolcirc-equations}} d$.
\end{lemma}
\begin{proof}
  The idea is that one may put circuits in a canonical form,
  so that then it is straightforward to eliminate all the unsafe paths by
  iteratively applying $\inlinefig{axioms/copy-mul} = \gidentity{}$ as a
  rewrite rule. See Appendix~\ref{appendix:polynomials} for the full proof.
\end{proof}

By virtue of Lemma~\ref{prop:Rsafenormalform}, in order to show that
$\R$ yields a well-defined operator on the whole of $\BoolCirc$,
it suffices to show
that it is well-defined on safe circuits. To do so, the following is the key
intermediate lemma.

\begin{lemma}
  \label{lemma:safety-implies-a-equivalence}
  If two circuits $c$ and $d$ are safe and
  $c \eqmod{\eqref{equation:boolcirc-equations}} d$,
  then $c \eqmod{\Axioms} d$.
\end{lemma}

\begin{proof} We give an overview of the structure of the argument and refer to
  Appendix~\ref{appendix:polynomials} for the full details.
  The proof relies on the polynomial interpretation of circuits,
  $\ToPoly{\cdot}$.
  Under this interpretation,
  $\ToPoly{\inlinefig{axioms/copy-mul}} = \langle x^2 \rangle$ and
  $\ToPoly{\gidentity{}} = \langle x \rangle$. Thus $\inlinefig{axioms/copy-mul}
  =\gidentity{}$ is not sound under this interpretation, because $\langle x^2
  \rangle \neq\langle x \rangle$.
  On the other hand, we know that $c \eqmod{\Axioms} d$ if and only if
  $\ToPoly{c} = \ToPoly{d}$ by \autoref{proposition:polycirc-presents-polyz}.
  By definition, this is the same as saying that $c \eqmod{\eqref{equation:boolcirc-equations}} d$ if and only if $\ToPoly{c} = \ToPoly{d}$ modulo the equation $x^2 = x$.

Now, one can prove that, for any safe circuit $c$, $\ToPoly{c}$ does not contain
  any squared term (Lemma~\ref{lemma:no-squared-terms}). Thus, coming to  $c$
  and $d$ as in the statement of the lemma, because $c$ and $d$ are such that $c
  \eqmod{\eqref{equation:boolcirc-equations}} d$, it follows that $\ToPoly{c} =
  \ToPoly{d}$ modulo the equation $x^2 = x$. Because $c$ and $d$ are safe, they
  do not contain any squared terms, and thus $\ToPoly{c} = \ToPoly{d}$. By
  completeness of $\Axioms$ with respect to $\ToPoly{\cdot}$, we conclude that
  $c \eqmod{\Axioms} d$.
\end{proof}

We can now conclude that

\begin{proposition}
  \label{theorem:r-well-defined}
  $\R$ is well-defined on safe circuits modulo
  \eqref{equation:boolcirc-equations}, that is, for $c$ and $d$ safe,
  if $c \eqmod{\eqref{equation:boolcirc-equations}} d$ then $\R[c]
  \eqmod{\eqref{equation:boolcirc-equations}} \R[d]$.
\end{proposition}

\begin{proof}
  Suppose we have two safe circuits $c, d$ such that $c \eqmod{\eqref{equation:boolcirc-equations}} d$.
  By \autoref{lemma:safety-implies-a-equivalence} we know that
  $c \eqmod{\Axioms} d$,
  and therefore we have
  $\R[c] \eqmod{\Axioms} \R[d]$ by \autoref{lemma:r-well-defined-on-axioms}.
  Finally, because $\Axioms \subseteq \eqref{equation:boolcirc-equations}$, we have that
  $\R[c] \eqmod{\eqref{equation:boolcirc-equations}} \R[d]$.
\end{proof}

We can now define an operator $R$ of boolean circuits which computes the reverse
derivative.

\begin{definition}
  Let $f$ be a morphism of $\BoolCirc$, i.e. a
  \eqref{equation:boolcirc-equations}-equivalence class of boolean circuits. Let
  $c$ be a safe circuit in such class, which exists by
  Lemma~\ref{prop:Rsafenormalform}. Define $R[f]$ as the
  \eqref{equation:boolcirc-equations}-equivalence class of $\R[c]$.
  Since $c$ is safe, we know that $R$ is well defined thanks to
  Proposition~\ref{theorem:r-well-defined}.
\end{definition}

Note that this definition is a minor abuse of notation, because $R$
does \textit{not} make \BoolCirc{} a reverse derivative category.
This is because the safety condition is not compositional, and thus cannot
satisfy axiom RD.5.
Nevertheless, we are still able to use $R$ to learn the parameters of boolean
functions, as we demonstrate in the following sections.

\section{\RDA{}}
\label{section:reverse-derivative-ascent}

\subsection{\RDA{} Algorithm}

We now introduce our machine learning algorithm,
\textdef{reverse derivative ascent}. The definition refers to the category
$\BoolCirc$, as boolean circuits are our motivating example. However, our
formulation makes sense in any reverse differential category.

We proceed in two parts: the inner `step' of the algorithm, which we call
\rdaStep{}, and the outer `iteration' of \rdaStep{}, which is \rda{}.

\begin{definition}
  Let $f \from p + a \to b$ be a boolean circuit in $\BoolCirc$, thus computing a parametrised
  boolean function with $p$ parameters.
  We define $\rdaStep{}_f$ as
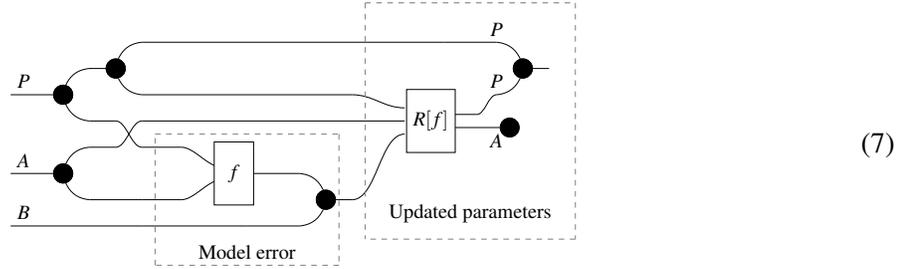
  \begin{equation}
    \label{equation:rda-step}
    \biginlinefig{rda-step}{2cm}
  \end{equation}
\end{definition}

  $\rdaStep{}_f$ represents a single iteration of \rda{}.
  Its function is to compute a new parameter vector $\theta'$ for a single
  labelled dataset example $(x, y) \in (\Z^a, \Z^b)$.
  We highlight two important parts of \rdaStep{}.
  First, it computes the \textit{model error} $\delta_y := f(\theta, x) + y$:
  the difference in model prediction to true label.
  Secondly, it uses $R[f]$ to compute a \textit{change in parameters}
  $\delta_\theta := R[f](\theta, x, \delta_y) \pi_0$ \footnote{
    Following \cite{cockett_reverse_2019}, we write composition left-to-right to
    mimic diagrammatic order.
  } such that
  $f(\theta + \delta_\theta, -)$ will more closely approximate the example datum.

Of course, we would like to update our parameters multiple times: this is the
\textit{ascent} part of reverse derivative ascent.
In Haskell \rda{} is simply the \texttt{scanl} operation over \rdaStep{},
but we define \rda{} as a circuit to emphasize its generality as a morphism
of a reverse derivative category.

\begin{definition}
  Let $n \in \mathbb{N}$,
  and let $(x_i, y_i) \in (\Z^a, \Z^b)$,
  denote a sequence of examples with $0 < i \leq n$.
  $\rda{}_f$ is defined as
  \begin{equation}
    \label{equation:rda-iteration}
    \biginlinefig{rda-iteration}{2.5cm}
  \end{equation}
\end{definition}

\begin{remark}
  In general, there is no need for elements $(x_i, y_i)$ to be in direct
  correspondence with elements of the dataset. Commonly, the sequence of
  examples `shown' to the algorithm will be shuffled with repetitions
  \cite[6.1]{ruder_overview_2017}.
\end{remark}

\subsection{Empirical Results}
\label{subsection:empirical-results}

We now show empirical results of our method (\autoref{table:empirical-results}),
which suggest that our algorithm is genuinely able to learn useful functions
from real-world data.
The full source code for running these experiments
is available at \url{http://catgrad.com/p/reverse-derivative-ascent}.
We begin with a brief discussion of our implementation.

\subsubsection{Implementation Details}
\label{subsubsection:implementation-details}

The purpose of our implementation is to specify and evaluate circuits as
machine learning models.
A boolean circuit $a \to b$ is represented by a term of the datatype
\texttt{a :-> b}; more complex circuits are built by composition and tensoring
from the primitives of \eqref{equation:circuit-generators}.
Note that, in our implementation, such primitives already come interpreted as the
corresponding boolean functions, exploiting the isomorphism between \BoolCirc{}
and \BoolFun{} (\autoref{proposition:circ-fun-iso}).
This is just a presentation choice, which spares us the need to define a
separate syntax and interpreter.
In the future, we plan to enhance the flexibility of our tool by making these
two components distinct.

In fact, our datatype \texttt{:->} is a \textit{pair} of a circuit and its
reverse derivative: constructing a circuit simultaneously
constructs its reverse derivative precisely as in
\eqref{equation:r-definition-composition}.\footnote{
  Interestingly, this pairing is the way in which the reverse derivative
  construction can be made functorial.
  See \cite[Proposition 31]{cockett_reverse_2019} for details.
}
In this way, the reverse derivative is built up compositionally from smaller
parts, and therefore to compute the reverse derivative we need only to extract
the second element of the pair, for which we provide the \texttt{rdiff}
function.

As we have seen in
Section \ref{section:graphical-boolean-reverse-derivative}, composing reverse
derivatives in this way is only valid for \textit{safe} circuits.
This prototype version of the code does not implement the procedure described in
\autoref{appendix:polynomials} to extract a safe circuit,
and so we provide a second method to compute reverse derivatives:
the brute-force \texttt{rdiffB} function (as described in \autoref{example:boolfun-rd}).
Note that this method can be applied even to unsafe circuits, but is
significantly less efficient compared to the compositional \texttt{rdiff} as
defined in \autoref{def:Rtilde}.
For example, in our experiment code we consider the \texttt{eval} model--an instance of
which we show in \autoref{example:eval-circuit}.
For $a$-dimensional input, the model has $2^a$ parameters, and so computing
\texttt{rdiffB eval} requires running \texttt{eval} an exponential number of times.
By comparison, $R[\texttt{eval}]$ (as computed by \texttt{rdiff eval}) is a
circuit whose size is within a constant factor of $\texttt{eval}$, and whose
result needs to be computed just once.

To showcase the difference, in the two experiments which follow, we use
\texttt{rdiff} for compositionally for the Iris model--since it is safe--but use
\texttt{rdiffB} for the MNIST model.
In the latter case, the number of parameters is equal to the number of inputs,
so this method is not too computationally demanding.

We discuss further avenues for improvement to our prototype implementation in
Section \ref{section:discussion}.

\subsubsection{Iris Dataset \& Model}
\label{subsubsection:iris-model}

\begin{table}
  \caption{Empirical Results}
  \label{table:empirical-results}
  \centering
  \begin{tabular}{lcll}
    \toprule
    Dataset   & Model      & Label Encoding  & Accuracy \% \\
    \midrule
    Iris (2-class)  & eval $: 16 + 4 \to 1$   & binary  & 98.0\%  \\
    Iris (2-class)  & eval $: 16 + 4 \to 2$   & one-hot & 98.0\%  \\
    \midrule
    Iris            & eval $: 16 + 4 \to 2$   & binary  & 73.3\%   \\
    Iris            & eval $: 16 + 4 \to 3$   & one-hot & 73.3\%   \\
    \midrule
    MNIST (2 class) & pseudoLinear $: 784 + 784 \to 1$ & binary  & 99.2\%   \\
    \bottomrule
  \end{tabular}
\end{table}

The Iris dataset \cite{Dua:2019} is a simple example of a classification
problem,
and is
frequently used for pedagogical purposes, e.g.
in \cite{duda_pattern_2000}.
It consists of 150 labelled examples of three types of iris flower.
Each example consists of four measurements of the flower petal and sepal sizes,
so we have the dataset of examples
$(\tilde{x}, \tilde{y}) \in (\mathbb{R}^4, \{\mathrm{Setosa}, \mathrm{Versicolor}, \mathrm{Virginica}\})$.

We run two experiments with this dataset, using our running example of the
\texttt{eval} model.
We first tackle the simpler problem of
the sub-dataset consisting of the labeled examples for classes Setosa and
Versicolor, which we call `Iris (2-class)' in \autoref{table:empirical-results},
and then show results for the full 3-class problem.
We also run two variations of each of these two experiments, corresponding to
different ways of encoding the labels: the 3-bit one-hot encoding\footnote{
  \texttt{one-hot} refers to the standard
  practice of encoding the $i$th class label of $n$ total as a vector with zero
  entries except for the $i$th.
  See e.g., \cite[section 3.3]{bishop_pattern_2006}
}, and the encoding of labels as binary numbers.
In all experiments, we preprocess this data by \textit{normalizing} and \textit{rounding} each
feature $\tilde{x}_i \in \mathbb{R}$ into a single bit $x_i \in \Z$.\footnote{
  This is essentially throwing away as much of the information of the dataset as
  possible: we map each feature to a simple `high' or `low' value
}
For the $n$-class problem, this gives us a
dataset of examples $(x, y) \in (\Z^4, \Z^n)$ for the one-hot encoding, and
$(x, y) \in (\Z^4, \Z^{\lceil \mathrm{log}_2(n) \rceil})$ for the binary encoding.

\subsubsection{MNIST Dataset \& Model}

MNIST \cite{Lecun98gradient-basedlearning} is an image classification dataset
widely used as a benchmark in machine learning
(see e.g., \cite{courbariaux_binaryconnect_nodate}).
It consists of 60000 examples of images of handwritten numeric digits (0 to 9),
with each image consisting of $28 \times 28$ greyscale pixels encoded as bytes.
The dataset therefore consists of examples
$(\tilde{x}, \tilde{y}) \in (\{0..255\}^{28 \times 28}, \{0..9\})$.

We do not tackle the full 10-class problem, but leave it for future work.
Instead, we restrict ourselves to the subset of classes $\{0, 1\}$.
While this means we cannot compare our method to the state of the art on this
benchmark, we believe it demonstrates that our method is indeed capable of
learning.
As in the Iris data, we also \textit{binarize} the pixels of the dataset by
normalisation and rounding, to give our `binarized' dataset of examples
$(x, y) \in (\Z^{28 \times 28}, \Z)$

Clearly the dimensionality of this problem is too large to use \texttt{eval}, so
we instead use a model
$\texttt{pseudoLinear} \from (28 \times 28) + (28 \times 28) \to 1$,
so named because its structure is loosely
inspired by the linear layers of neural networks.
We give only a brief informal description of this model here (for technical
details, see the experiment code we release with this paper \footnote{
  \url{https://github.com/statusfailed/act-2020-experiments}
})

Essentially, the model learns a `feature mask', which is simply a bitmap image
that is pointwise multiplied with the input. If the resulting bitvector has
fewer than $25\%$ as many $1$ bits as the mask, the model returns $1$.
The intuition is that the model should learn the `average' handwritten $0$
digit, and compare it with inputs.
In the two-class case this is a fair assumption, since images of $1$ and $0$
are typically very different, but it is unlikely to generalise well.

\subsubsection{Discussion of Results}

From \autoref{table:empirical-results}, we can see that the \texttt{eval} model
is able to learn a near-perfect classifier for the 2-class problem, but fares poorly
on the full problem. This is because our preprocessing essentially limits the
model to fixed, axis-aligned decision boundaries. Since the Setosa and
Versicolor classes are clearly separable when plotted, this works well, but the
Versicolor and Virginica classes are not.
We also note that because \texttt{eval} is essentially a lookup table, the
label encoding has no effect on model accuracy. This suggests that
\texttt{eval} may be useful as an `output unit' in larger models.

Finally, we note that our MNIST model, while only classifying a subset of the
full problem, returns fairly good results. To make an apples-to-oranges
comparison, the approach of \cite{courbariaux_binaryconnect_nodate} gives a
similar accuracy of $99.13\%$.
However, such a comparison is to be taken with a grain of salt: the full MNIST
problem is of course much more difficult than the version we tackle here.

\section{Discussion and Future Work}
\label{section:discussion}

In this paper, we saw how the categorical axiomatisation of reverse derivative
can be used to define a general `gradient' based algorithm for machine learning.
Further, we showed how our algorithm can be used to learn parameters of a novel
model class: boolean circuits.
However, there are many opportunities for future work, which we broadly classify
into two parts.

\paragraph{Empirical Work}
The first task is to discover principles for building effective
parametrised circuit models.
While a number of compositional building blocks for neural network
models have been discovered and studied, the same is not true for
parametrised boolean circuits.
One exciting challenge is to understand whether neural network
architectures can be translated to the setting of circuits

Furthermore, although our empirical results show our algorithm is certainly able
to learn parameters from data, a new machine learning method would typically be
expected to show results on the \textit{full} MNIST problem, as well as other
image processing benchmarks like CIFAR \cite{krizhevsky_learning_2009}.
Therefore, some empirical study of circuit architectures with respect to these
benchmarks will have to be undertaken.

As mentioned in Section \ref{subsubsection:implementation-details},
another important point is to enhance our implementation.
First, we intend to clearly separate between boolean circuits and their
interpretations as boolean functions, so that other semantic interpretations are
possible.
Second, we plan to implement our procedure to turn a circuit into its safe
equivalent, and study its complexity.

\paragraph{Theoretical Work}
One avenue for theoretical work is to demonstrate the use of \RDA{} on
categories other than boolean circuits.
For example, when interpreted in
the category of natural numbers and morphisms $a \to b$
the smooth maps $\mathbb{R}^a \to \mathbb{R}^b$,
our method is similar to stochastic gradient descent (SGD)
\cite{ruder_overview_2017}, with the following differences.
Firstly, computing the model error (\eqref{equation:rda-step}) means computing
the \textit{difference} between true label and model prediction, but in $\Z$
this coincides with addition because elements are self-inverse.
Secondly, SGD has a notion of \textit{learning rate}: a constant multiplied by
the parameter change which prevents the algorithm `overshooting' the optimal
parameter value.
Thirdly, we have no explicit \textit{loss function}, which is important to
discover the conditions under which guarantees of convergence exist.
By comparison, several different guarantees of convergence are known for
different variants of gradient descent as used in neural networks (see e.g.,
\cite[p. 2]{ruder_overview_2017}.), although in some cases tweaks such as slowly
decreasing the learning rate are required to make such guarantees, and prevent
oscillation around local minima.

Another setting of interest is boolean circuits with notions
of \textit{feedback}--something which has already received attention in the
literature \cite{sprunger_differential_2019}
\cite{sprunger_differentiable_2019}.
Characterising these
differences between settings may help to understand gradient methods in a more
general light.

It will also be important to relate our work to existing category theoretic
views of gradient-based methods such as \cite{gavranovic_learning_nodate}
\cite{fong_backprop_2019}.
In particular, we believe our method is a special case of
\cite{fong_backprop_2019}. Concretely, we note that for two parametrised boolean
functions $f \from p + a \to b, g \from q + b \to c$, taking the reverse
derivative of their `parametrised composition'
$(\mathrm{id} \times f) g \from q + p + a \to c$ is exactly the composite
update-request morphism from their formalism.

\paragraph{Acknowledgements}
We are grateful to the reviewers for their insightful comments and remarks.
We would also like to thank David Sprunger and Liviu Pirvan for several helpful
discussions.

\bibliographystyle{eptcs}
\bibliography{main}

\appendix
\section{Polynomial Interpretation of Boolean Circuits}
\label{appendix:polynomials}

In \autoref{subsection:r-well-defined-proof} we used the interpretation of circuits with
axioms $\Axioms{}$ as morphisms of \PolyZ{}.
We now make this interpretation precise.
We will discuss two results for these polynomial circuits:
soundness and completness of their interpretation $\ToPoly{\cdot}$,
and a canonical form.

\subsection{Soundness and Completeness}

To show the existence of an isomorphism between $\PolyCirc$ and $\PolyZ$,
we will show that both categories' hom-sets have the structure of the free
module over the polynomial ring. For this, we must recall the definition of a
free module:

\begin{definition}
  Following \cite[p. 170]{jacobson_basic_2012},
  let $S$ be a ring. The free module $S^b$ is the cartesian product of $b$
  elements of $S$, i.e.
  $S^b = \langle p_1, p_2, ..., p_b \rangle$,
  with addition defined pointwise,
  $ \langle p_1, p_2, ... p_b \rangle + \langle q_1, q_2, ... q_b \rangle
      = \langle p_1 + q_1, p_2 + q_2, ... p_b + q_b \rangle $
  a zero element $ \FZ = \langle 0, 0, ..., 0 \rangle $
  and scalar multiplication
  $ s \langle p_1, p_2, ..., p_b \rangle
      = \langle s p_1, s p_2, ... s p_b \rangle $
\end{definition}

It is clear that the hom-sets of $\PolyZ$ have this structure

\begin{proposition}
  \label{proposition:polyz-hom-free-module}
  Hom-sets $\PolyZ(a, b)$ have the structure of the free module
  $S^b$ with $S$ the polynomial ring $S = \Z[x_1, ..., x_a]$.
\end{proposition}
\begin{proof}
  Immediate from the definition of \PolyZ
\end{proof}

Furthermore, hom-sets of $\PolyCirc$ also have this structure.
This implies the existence of a module
isomorphism between the hom-sets of $\PolyCirc$ and $\PolyZ$
which is the basis for the functor
$\ToPoly{\cdot}$.
We begin, however, with some special case examples.

\begin{example}
  \label{example:hom-0-1-ring}
  The hom-set $\PolyCirc(0, 1)$ has the structure of the ring \Z{},
  with every circuit $c$ equal to $\gzero$ or $\gone$.
\end{example}

\begin{example}
  Each hom-set $\PolyCirc(a, 1)$ has the structure of the polynomial ring
  $\Z[x_1, ..., x_a]$,
  with indeterminates $x_1 \ldots x_a$
  given by the projections $\pi_1 \ldots \pi_a$
\end{example}

\begin{proposition}
  \label{proposition:polycirc-hom-free-module}
  Hom-sets $\PolyCirc(a, b)$ have the structure of the free module
  $\Z[x_1, \ldots, x_a]^b$.
\end{proposition}

\begin{proof}
  For morphisms $f, g \from a \to b$,
  put addition $f + g = \biginlinefig{addition}{0.5cm}$
  and multiplication \\ $f * g = \biginlinefig{multiplication}{0.5cm}$,
  with the zero element defined as $\FZ{} = \inlinefig{zero-morphism}$.
  one can verify graphically using equations $\Axioms{}$ that the module axioms
  hold.
  \footnote{
    We take \textit{scalar} multiplication of $f \from a \to b$ by $g \from a
    \to 1$ as the morphism
    $f * (g \Delta^*)$,
    where $\Delta^*$ is the unique $1 \to b$ morphism formed by tensor and
    composition of the diagonal map and identity.
  }
  If we define the family of $b$ morphisms $e_i :=
  \biginlinefig{basis-vector}{0.6cm}, 0 < i \le b$,
  we can see that it forms a base: each of the generators of
  \autoref{equation:circuit-generators} can be constructed through addition and
  scalar multiplication of morphisms $e_i$ and $\FZ$.
\end{proof}

We are now ready to give the proof of Proposition~\ref{prop:polyz-iso-polycirc}.

\begin{proof}[Proof of Proposition~\ref{prop:polyz-iso-polycirc}]
  By \autoref{proposition:polyz-hom-free-module}
  and \autoref{proposition:polycirc-hom-free-module},
  there is a module isomorphism between $\PolyZ(a, b)$ and $\PolyCirc(a,b)$.
  Further, because the identity-on-objects functor $\ToPoly{\cdot}$ is defined
  in terms of this bijection, it is a full and faithful functor,
  and so $\PolyZ \cong \PolyCirc$.
\end{proof}


\subsection{Canonical Form}

We now give a canonical form for morphisms of \PolyCirc{}.
This canonical form essentially isolates all occurrences of the axiom
$\inlinefig{axioms/copy-mul} = \gidentity{}$,
which we can then use to show that all boolean circuits have a safe equivalent
in the proof of \autoref{prop:Rsafenormalform}.

\begin{definition}
  \label{definition:algebraic-normal-form}
  We say a circuit $f \from a \to b$ is in canonical form if it can be written
  as \\
  $\langle \ToCanonical{p_1}, \ToCanonical{p_2}, \ldots, \ToCanonical{p_b} \rangle$,
  where $\langle p_1, \ldots, p_b \rangle = \ToPoly{f}$,
  and $\ToCanonical{-}$ is defined on polynomials as follows.

  Denote by $\pow{n}$ the morphism defined inductively as $\pow{0} = \gdiscard{}
  \gzero{}$ and $\pow{n} = \biginlinefig{pow-n}{0.4cm}$,
  and let $x_i^k$ be an arbitrary indeterminate raised to a power $k \in
  \mathbb{N}$.
  We define $\ToCanonical{x_i^k} := \biginlinefig{pi-pow}{0.5cm}$,
  and note that this is consistent with $\ToPoly{\cdot}$ in the sense that
  $\pi_i * \stackrel{k}{\ldots} * \pi_i \eqmod{\Axioms} \ToCanonical{x_i^k}$.

  We now define $\ToCanonical{\cdot}$ inductively on polynomials.
  Either $p$ is a constant, in which case
  $\ToCanonical{0} = \inlinefig{zero-morphism}$
  and $\ToCanonical{1} = \inlinefig{one-morphism}$,
  or $p = m + p'$ and we have $\ToCanonical{m} + \ToCanonical{p'}$,
  where $m$ is a monomial and $p'$ a polynomial.
  A monomial is a product of \textit{distinct} indeterminates raised to powers
  $x_i^{k_i}, x_j^{k_j}, ...$,
  and its canonical form is therefore
  $\ToCanonical{x_i^{k_i}} * \ToCanonical{x_j^{k_j}} * ...$.
\end{definition}

\begin{remark}
  Intuitively, the canonical form can be pictured as
  $ \biginlinefig{canonical-form}{0.6cm} $
\end{remark}

\begin{example}
  Continuing with our running example, we note that the
  $\texttt{eval} \from 2 + 1 \to 1$ circuit can be written as the polynomial
  $x_1 + x_1 x_3 + x_2 x_3$ (where parameters are $x_1$ and $x_2$), and so its
  safe canonical form can be written as
  $$\biginlinefig{eval-normal-form}{1.8cm}.$$
  To see it is equivalent to \eqref{equation:eval-2-1-circuit}, we can apply the
  counit law $\inlinefig{axioms/copy-unit} = \gidentity{}$, and then use
  distributivity.
\end{example}

We are now ready to show the proof of \autoref{prop:Rsafenormalform}.

\begin{proof}[Proof of Lemma~\ref{prop:Rsafenormalform}]
  We show that for each circuit $c$ there is a safe circuit $d$ such that $c
  \eqmod{\eqref{equation:boolcirc-equations}} d$.
  We begin by noting that with equations \eqref{equation:boolcirc-equations},
  we have that $\pow{k} = \gidentity{}$, which can be seen by repeatedly applying the
  $\inlinefig{axioms/copy-mul} = \gidentity{}$ axiom.
  Using this, we can see that the canonical form of
  \autoref{definition:algebraic-normal-form} can be rewritten so that each
  $\pow{k}$ morphism becomes the identity.
  Finally, because each product $\biginlinefig{multiplication}{0.5cm}$ in the
  canonical form is of
  distinct indeterminates, the rewritten canonical form does not contain any more
  squared terms, and so is safe.
\end{proof}

To conclude, we show how the condition of safety essentially allows only those
circuits with interpretations as Zhegalkin polynomials \cite{zhegalkin_sur_1927},
as used in the proof of \autoref{lemma:safety-implies-a-equivalence}.

\begin{lemma}
  \label{lemma:no-squared-terms}
  If a circuit $f$ is safe, then the polynomial $\ToPoly{f}$ has only exponents
  in $\{0, 1\}$.
\end{lemma}

\begin{proof}
  The combinatorial condition (\autoref{def:safe}) requires that the inputs of each \gand connect to
  disjoint sets of inputs. Therefore, the resulting polynomial $\ToPoly{f}$ only
  contains multiplications of polynomials $p(\vec{x}_1) p(\vec{x}_2)$ of
  disjoint sets of variables, so there can be no squared terms in \ToPoly{f}.
\end{proof}

\end{document}